\documentclass[aps,pra,epsfig,twocolumn,showpacs,superscriptaddress]{revtex4}

\def\be{\begin{equation}}
\def\ee{\end{equation}}
\def\bea{\begin{eqnarray}}
\def\eea{\end{eqnarray}}
\def\bma{\begin{mathletters}}
\def\ema{\end{mathletters}}
\def\bi{\begin{itemize}}
\def\ei{\end{itemize}}

\def\C{\hbox{$\mit I$\kern-.7em$\mit C$}}

\begin{document}

\title{Distinguishability of maximally entangled states}

\author{Sibasish Ghosh}
\email{sibasish@imsc.res.in}
\affiliation{Institute of Mathematical Sciences, C. I. T. Campus, Taramani, Chennai 600113, India}

\author{Guruprasad Kar}
\email{gkar@imsc.res.in}
\affiliation{Institute of Mathematical Sciences, C. I. T. Campus, Taramani, Chennai 600113, India}

\author{Anirban Roy}
\email{anirb@imsc.res.in}
\affiliation{Institute of Mathematical Sciences, C. I. T. Campus, Taramani, Chennai 600113, India}

\author{Debasis Sarkar}
\email{debasis1x@yahoo.com}
\affiliation{Department of Applied Mathematics, University of Calcutta, 92 A. P. C. Road, Kolkata 700009, India}

\date{\today}

\begin{abstract}
In $2 \otimes 2$, more than two orthogonal Bell states with single
copy can never be discriminated with certainty if only local
operations and classical communication (LOCC) are allowed.  
More than $d$ orthogonal maximally entangled states in $d \otimes d$, which are in canonical form, used by Bennett et al. [{\it Phys. Rev. Lett.} {\bf 70} (1993) 1895], can never be discriminated with
certainty when a single copy of the states is provided.
Interestingly we show that all orthogonal maximally entangled
states , which are in canonical form, can be discriminated with certainty if and only if two
copies of each of the states are provided. The highly nontrivial problem of local discrimination of any $d$ or less no. of pairwise orthogonal maximally entangled states in $ d \otimes d$ (in single copy case), which are in canonical form, is also discussed here. 
\end{abstract}

\pacs{03.67.-a, 03.65.Bz, 03.67.Hk}

\maketitle

\section{Introduction}
In quantum mechanics, any set of orthogonal states can be
discriminated. But for multipartite system, local information of
the density matrices, and even local operations and classical
communication (LOCC) are not sufficient to distinguish among
orthogonal states.  Recently some interesting studies have shown
that pairwise orthogonal multipartite states cannot always be
discriminated with certainty in a single copy case if only local
operations and classical communication (LOCC) are allowed
\cite{bennett99, hardy00, hardy02, ghosh01}. But any two
multipartite orthogonal states can always be distinguished with
certainty by LOCC, and, in general, $d$ pairwise orthogonal multipartite states
can be perfectly discriminated by LOCC if $(d-1)$ copies of each state are
provided \cite{hardy00}. But there are sets of pairwise orthogonal
states that can be discriminated with less than $(d-1)$ copies.
One such example is that if \emph{two copies} of a state are
provided which is known to be one of the four pairwise orthogonal
Bell states
\begin{equation}
\begin{array}{lcr}
\label{bellstates} \left| {\Phi}^{\pm}\right\rangle &=&
\frac{1}{\sqrt{2}}\left( \left| 00\right\rangle \pm \left|
11\right\rangle \right),\\
\left| {\Psi}^{\pm}\right\rangle &=& \frac{1}{\sqrt{2}}\left(
\left| 01\right\rangle \pm \left| 10\right\rangle \right),\\
\end{array}
\end{equation}
one can discriminate between them using LOCC only \cite{hardy00}.
In a recent paper Ghosh \emph{et. al.} \cite{ghosh01} have shown,
using some properties of entanglement measure, that more than two
orthogonal Bell states cannot be discriminated with certainty if a
single copy is provided. 

In this paper, we consider the problem of reliable local distinguishability of pairwise orthogonal maximally entangled states in $d \otimes d$, each of which are in canonical form, given by equation (\ref{psinm}), below. Let let $\{|0\rangle, |1\rangle, \ldots, |d - 1\rangle\}$ be the standard orthonormal basis of a $d$-dimensional Hilbert space. A {\it maximally entangled state} in $d \otimes d$ is defined to be a pure state for which both the reduced density matrices are equal to the maximally mixed state $\frac{1}{d}I$, in a $d$-dimensional Hilbert space. In the above-mentioned standard basis, $d^2$ no. of pairwise orthogonal maximally entangled states can be written as:
\begin{equation}
\label{psinm} 
\left|{\Psi}_{nm}^{(d)}\right\rangle =
\frac{1}{\sqrt{d}} \sum_{j = 0}^{d - 1}~ {\rm exp}\left[\frac{2
\pi i j n}{d}\right] \left|j\right\rangle \otimes \left|\left(j +
m\right)~ {\rm mod}~ d\right\rangle, 
\end{equation} 
for $ n, m = 0, 1, \ldots, d - 1$. 

Here we show in section II that more than $d$ pairwise orthogonal maximally entangled
states in $d \otimes d$, all taken from the set given in equation (\ref{psinm}), can never be perfectly discriminated by
LOCC in a single copy case. As mentioned above,
in  $2 \otimes 2$, the set of four (or three) orthogonal Bell states can be
discriminated with certainty, using LOCC only, if at least two copies of
each state are given. Interestingly this is universal, {\it i.e.},
(we show here in section III that) any number of mutually orthogonal
maximally entangled states in $d \otimes d$, all taken from the set given in equation (\ref{psinm}), can be discriminated
by LOCC only, if two copies of the states are provided. This is
definitely surprising as one would be inclined to think that the
minimum number of copies needed for discrimination would be an
increasing function of the dimension $d$. Next, in section IV, we discuss the problem of reliable local discrimination of any $d$ or less no. of pairwise orthogonal maximally entangled states in $d \otimes d$, taken from the set given by equation (\ref{psinm}), in the single copy case \cite{fan03}, based on a particular type of 1--way LOCC, namely, teleportation. In section V, we discuss a necessary condition for reliable distinguishability via LOCC. Finally, in section VI, we draw the conclusion.

Before going into the main results, let us discuss briefly some properties of the maximally entangled states, given by equation (\ref{psinm}). It is easy to verify that 
\begin{equation}
\label{urelation}
\begin{array}{lcr}
\left(U_{nm}^{(d)} \otimes I\right) \left|{\Psi}_{00}^{(d)}\right\rangle &=& \left|{\Psi}_{nm}^{(d)}\right\rangle,\\ \\
\left(I \otimes V_{nm}^{(d)}\right) \left|{\Psi}_{00}^{(d)}\right\rangle &=& \left|{\Psi}_{nm}^{(d)}\right\rangle,
\end{array}
\end{equation}
where 
\begin{equation}
\label{udefine}
U_{nm}^{(d)} = \sum_{j = 0}^{d - 1}~ {\rm exp} \left[\frac{2 \pi i j n}{d}\right]~ |j\rangle\langle(j + m)~ {\rm mod}~ d|,
\end{equation}
and 
\begin{equation}
\label{vdefine}
V_{nm}^{(d)} = \left(U_{((d - n)~ {\rm mod}~ d)m}^{(d)}\right)^{\dagger},
\end{equation}  
for $n, m = 0, 1, \ldots d - 1$. It should be noted here that:

{\it orthogonality of any two maximally entangled states $\left|{\Psi}_1^{(d)}\right\rangle = \left(I \otimes V_1^{(d)}\right)\left|{\Psi}_{00}^{(d)}\right\rangle$ and $\left|{\Psi}_2^{(d)}\right\rangle = \left(I \otimes V_2^{(d)}\right)\left|{\Psi}_{00}^{(d)}\right\rangle$ of $d \otimes d$, which are not necessarily of the form of equation (\ref{psinm}), is equivalent to the `trace-orthogonality' condition ${\rm Tr} \left[{\left(V_1^{(d)}\right)}^{\dagger}V_2^{(d)}\right] = 0$.}

 From equations (\ref{urelation}) - (\ref{vdefine}), one gets that
\begin{equation}
\label{newrelation}
\begin{array}{lcc}
\left|{\Psi}_{nm}^{(d)}\right\rangle &=& \left(U_{nm}^{(d)}\left(U_{n^{\prime}m^{\prime}}^{(d)}\right)^{\dagger} \otimes I\right) \left|{\Psi}_{n^{\prime}m^{\prime}}^{(d)}\right\rangle, \\ \\
\left|{\Psi}_{nm}^{(d)}\right\rangle &=& \left(I \otimes V_{nm}^{(d)}\left(V_{n^{\prime}m^{\prime}}^{(d)}\right)^{\dagger}\right) \left|{\Psi}_{n^{\prime}m^{\prime}}^{(d)}\right\rangle,
\end{array}
\end{equation}
where
\begin{widetext}
\begin{equation}
\label{uvexpression}
\begin{array}{lcc}
U_{nm}^{(d)}\left(U_{n^{\prime}m^{\prime}}^{(d)}\right)^{\dagger} &=&~ {\rm exp} \left[\frac{2 \pi i (m^{\prime} - m) n^{\prime}}{d}\right] U_{((d + n - n^{\prime})~ {\rm mod}~ d)((d + m - m^{\prime})~ {\rm mod}~ d)}^{(d)}, \\ \\
V_{nm}^{(d)}\left(V_{n^{\prime}m^{\prime}}^{(d)}\right)^{\dagger} &=&~ {\rm exp} \left[\frac{2 \pi i (n^{\prime} - n) m^{\prime}}{d}\right] V_{((d + n - n^{\prime})~ {\rm mod}~ d)((d + m - m^{\prime})~ {\rm mod}~ d)}^{(d)},
\end{array}
\end{equation} 
$n, m, n^{\prime}, m^{\prime} = 0, 1, \ldots, d - 1$. 
\end{widetext}

In the present paper, we shall repeatedly use teleportation of some state $\left|\phi^{(d)}\right\rangle$ of a $d$-dimensional Hilbert space, via some shared maximally entangled state $\left|{\Psi}_{max}^{(d)}\right\rangle$ of $d \otimes d$ (which is not necessarily of the form given in (\ref{psinm})), using complete von Neumann measurement in maximally entangled basis (which are not necessarily of the form given in (\ref{psinm})) $\left\{\left|\Phi_i^{(d)}\right\rangle~ :~ i = 1, 2, \ldots d^2\right\}$ of $d \otimes d$ (and then using corresponding unitary operations). Thus, for the shared channel state $\left|{\Psi}_{00}^{(d)}\right\rangle$, if $\left|\Phi_i^{(d)}\right\rangle = \left(U_i \otimes I\right)\left|{\Psi}_{00}^{(d)}\right\rangle$ clicks in the measurement of Alice ($U_i$ being an unitary operator), in order to have {\it exact} teleportation (as described in \cite{bennett93}), Bob will have to apply this unitary operator $U_i$ on his system, so that, the output state at Bob's side will be $\left|\phi^{(d)}\right\rangle$ \cite{braunstein}. Let us denote this (exact) teleportation protocol as ${\cal P}\left(\left|{\Psi}_{00}^{(d)}\right\rangle; U_1, U_2, \ldots, U_{d^2}\right)$. On the other hand, if the shared channel state is $\left|{\Psi}_{max}^{(d)}\right\rangle = (I \otimes V)\left|{\Psi}_{00}^{(d)}\right\rangle$ ($V$ being an unitary operator), and if $\left|\Phi_i^{(d)}\right\rangle = \left(U_i \otimes I\right)\left|{\Psi}_{00}^{(d)}\right\rangle$ clicks in the measurement of Alice, then application of the unitary operator $U_i$ (by Bob) will give rise to the state $\left(U_iVU_i^{\dagger}\right)\left|\phi^{(d)}\right\rangle$. Thus, in this case, for the input state $\left|\phi^{(d)}\right\rangle$, the final output state ({\it i.e.}, for all $i$) will be a pure state if and only if all $\left(U_iVU_i^{\dagger}\right)\left|\phi^{(d)}\right\rangle$'s represent the same state (upto some phases). Now for $U_i = U_{nm}^{(d)}$ (given by equation (\ref{udefine})) and $V = V_{n^{\prime}m^{\prime}}^{(d)}$ (given by equation (\ref{vdefine})), we have $U_i V U_i^{\dagger} =  U_{nm}^{(d)} V_{n^{\prime}m^{\prime}}^{(d)}  \left(U_{nm}^{(d)}\right)^{\dagger} =~ {\rm exp}~ \left[\frac{2 \pi i \{n(m + m^{\prime}) - n^{\prime}(m - m^{\prime})\}}{d}\right] V_{n^{\prime}m^{\prime}}^{(d)}$. Thus, in this particular example, $P\left[\left(U_iVU_i^{\dagger}\right)\left|\phi^{(d)}\right\rangle\right] = P\left[V\left|\phi^{(d)}\right\rangle\right]$ for all $i$, where $P[.]$ stands for projector on the vector within the square bracket. We, therefore, see that if the shared channel state between Alice and Bob is $\left|{\Psi}_{nm}^{(d)}\right\rangle = \left(I \otimes V_{nm}^{(d)}\right) \left|{\Psi}_{00}^{(d)}\right\rangle$ (given in (\ref{urelation})), and if Alice does the von Neumann measurement in the basis $\left\{\left|{\Psi}_{n^{\prime}m^{\prime}}^{(d)}\right\rangle = \left(U_{n^{\prime}m^{\prime}}^{(d)} \otimes I\right) \left|{\Psi}_{00}^{(d)}\right\rangle : n^{\prime}, m^{\prime} = 0, 1, \ldots, d - 1\right\}$ (given in (\ref{urelation})), then for the input state $\left|{\phi}^{(d)}\right\rangle$, the output state will be $V_{nm}^{(d)}\left|{\phi}^{(d)}\right\rangle$ (upto a phase), after using the teleportation protocol  ${\cal P}\left(\left|{\Psi}_{00}^{(d)}\right\rangle; U_{00}^{(d)}, U_{01}^{(d)}, \ldots, U_{(d - 1) (d - 1)}^{(d)}\right)$. This teleportation protocol was given in \cite{bennett93}, and from now on, {\it we shall denote the protocol, in short, by ${\cal P}_{00}^{(d)}$}.

Next we consider the situation where the motivation is to have (exact) teleportation of an arbitrary state $\left|{\phi}^{(d)}\right\rangle$ from Alice's place to Bob's place, using the shared channel state as the maximally entangled state $\left|{\Omega}^{(d)}\right\rangle = (W \otimes I)\left|{\Psi}_{00}^{(d)}\right\rangle$ of $d \otimes d$ (instead of the channel state $\left|{\Psi}_{00}^{(d)}\right\rangle$), using complete von Neumann measurement (to be done by Alice) in the orthogonal basis of maximally entangled states $\left|\overrightarrow{{\Phi}_i^{(d)}}\right\rangle = \left(W_i \otimes I\right)\left|{\Omega}^{(d)}\right\rangle$ (where $W_i$'s are unitary operators for $i = 1, 2, \ldots d^2$), and using (to be done by Bob) respective unitary operators $T_i = W_iWW^*$ in order to get the twisted maximally entangled state $\left|\overleftarrow{{\Phi}_i^{(d)}}\right\rangle = \left(I \otimes T_i\right) \left|{\Omega}^{(d)}\right\rangle$. Thus (compared to ${\cal P}_{00}^{(d)}$), here we are using the (exact) teleportation protocol  ${\cal P}\left(\left|{\Omega}^{(d)}\right\rangle; W_1, W_2, \ldots, W_{d^2}\right)$. Thus, if $|\chi\rangle = (I \otimes V)\left|{\Omega}^{(d)}\right\rangle$ is the shared channel state in this situation, if $\left|{\phi}^{(d)}\right\rangle$ is the input state, and if $\left|\overrightarrow{{\Phi}_i^{(d)}}\right\rangle$ is the measurement outcome, the output state at Bob's side will be $\left(T_iVT_i^{\dagger}\right) \left|{\phi}^{(d)}\right\rangle$ \cite{braunstein}. And here, for the special case where we use the protocol 

\begin{widetext}
$${\cal P}\left(\left|{\Psi}_{nm}^{(d)}\right\rangle; U_{00}^{(d)} \left(U_{nm}^{(d)}\right)^{\dag}, U_{01}^{(d)}\left(U_{nm}^{(d)}\right)^{\dag}, \ldots, U_{(d - 1) (d - 1)}^{(d)}\left(U_{nm}^{(d)}\right)^{\dag}\right), $$ 
\end{widetext}

for the channel state $\left|{\Psi}_{n^{\prime \prime}m^{\prime \prime}}^{(d)}\right\rangle = \left(I \otimes {V}_{n^{\prime \prime}m^{\prime \prime}}^{(d)} \left({V}_{nm}^{(d)}\right)^{\dag}\right)\left|{\Psi}_{nm}^{(d)}\right\rangle$, the output states will be $V_{n^{\prime \prime}m^{\prime \prime}}^{(d)} \left(V_{nm}^{(d)}\right)^{\dagger} \left|{\phi}^{(d)}\right\rangle$, for all measurement outcomes. This shows that if $V_{n^{\prime \prime}m^{\prime \prime}}^{(d)} \left(V_{nm}^{(d)}\right)^{\dagger} \left|{\phi}^{(d)}\right\rangle$'s are pairwise orthogonal for all $n^{\prime \prime}, m^{\prime \prime} = 0, 1, \ldots, d - 1$, the states  $V_{n^{\prime \prime}m^{\prime \prime}}^{(d)} \left|{\phi}^{(d)}\right\rangle$'s are also pairwise orthogonal for all $n^{\prime \prime}, m^{\prime \prime} = 0, 1, \ldots, d - 1$, and vice versa. Thus, {\it with respect to this orthogonality requirement, whether we take the protocol 

\begin{widetext}
$${\cal P}\left(\left|{\Psi}_{nm}^{(d)}\right\rangle; U_{00}^{(d)} \left(U_{nm}^{(d)}\right)^{\dag}, U_{01}^{(d)}\left(U_{nm}^{(d)}\right)^{\dag}, \ldots, U_{(d - 1) (d - 1)}^{(d)}\left(U_{nm}^{(d)}\right)^{\dag}\right)~~ or~~ {\cal P}\left(\left|{\Psi}_{00}^{(d)}\right\rangle; U_{00}^{(d)}, U_{01}^{(d)}, \ldots, U_{(d - 1) (d - 1)}^{(d)}\right), $$ 
\end{widetext}

we will get the same result}. So in future, application of the teleportation protocol to test orthogonality of the output states, for different channel states, we will use ${\cal P}_{00}^{(d)}$.     


\vspace{0.4cm}
\section {Local indistinguishability of $(d + 1)$ no. of maximally entangled states}

We are now going to show that {\it no} $(d + 1)$ no. of pairwise orthogonal 
maximally entangled states in $d \otimes d$, all taken from the set given in (\ref{psinm}), can be reliably discriminated by LOCC, in the single copy case. For this, we shall use the notion of the relative entropy of entanglement $E_{r}(\sigma)$ for a bipartite quantum state \( \sigma \) \cite{vedral97}:

$$E_{r}(\sigma ) = \min _{\rho \in D}S(\sigma \parallel \rho ),$$

where \( D \) is the set of all separable states on the Hilbert
space on which \( \sigma  \) is defined, and \( S(\sigma \parallel
\rho )\equiv tr\{\sigma (\log _{2}\sigma -\log _{2}\rho )\} \) is
the relative entropy of \( \sigma  \) with respect to \( \rho  \).

Consider now the following four party state  

$$\rho ^{(d + 1)} = \frac{1}{(d + 1)}\sum
^{(d + 1)}_{i = 1}P\left[ \left|{\Psi}_{n_im_i}^{(d)}\right\rangle
_{AB}\left|{\Psi}_{n_im_i}^{(d)}\right\rangle _{CD}\right],$$

shared between Alice ($A$), Bob ($B$), Charlie ($C$) and Darlie ($D$) with
all four being at distant locations, where
$\left|{\Psi}_{n_im_i}^{(d)}\right\rangle$ (for $i = 1, 2, \ldots, d + 1$) are given any $(d + 1)$ no. of pairwise orthogonal maximally entangled states in $d \otimes d$, all taken from the set given in equation (\ref{psinm}). Consider now another four party state 

$$\rho^{(S)} = \frac{1}{d^2}\sum ^{d - 1}_{n, m = 0}P\left[ \left|
{\Psi}_{nm}^{(d)}\right\rangle _{AC}\left|
{\Psi}_{nm}^{(d)}\right\rangle _{BD}\right],$$

shared among $A$, $B$, $C$ and $D$, where $\left|{\Psi}_{nm}^{(d)}\right\rangle$'s are given by equation (\ref{psinm}). 
By construction, $\rho^{(S)}$ is separable across $AC : BD$ cut.
Let $E_{r}(\rho _{AC : BD}^{(d + 1)})$ be the
relative entropy of entanglement of the state $\rho ^{(d + 1)}$
in the $AC : BD$ cut. Then 

\begin{widetext}
$$E_{r}(\rho _{AC : BD}^{(d + 1)}) \leq S\left( \rho _{AC : BD}^{(d + 1)}\parallel \frac{1}{d^2}\sum ^{d - 1}_{n, m = 0}P\left[ \left|
{\Psi}_{nm}^{(d)}\right\rangle _{AC}\left|
{\Psi}_{nm}^{(d)}\right\rangle _{BD}\right] \right) = S\left( \rho _{AC : BD}^{(d + 1)}\parallel \rho^{(S)} \right) < \log_{2}d.$$
\end{widetext} 

But distillable entanglement is bounded above by \( E_{r} \) \cite{rains99, horo00}. Consequently the distillable
entanglement of \( \rho ^{(d + 1)} \), in the $AC : BD$ cut, is strictly less than $\log_{2}d$.

Suppose now that it is possible to discriminate (with certainty)
any $(d + 1)$ no. of pairwise orthogonal maximally entangled states $\left|{\Psi}_{n_im_i}^{(d)}\right\rangle$'s in $d \otimes d$, using only LOCC and
when only a single copy each of the state is provided. So if Alice, Bob, Charlie and
Darlie share the state \( \rho ^{(d + 1)} \), then Alice and Bob,
without meeting, would again be able to distill between Charlie and Darlie, \( \log_{2}d \) ebit of entanglement, 
by using this state-discrimination LOCC (together with possible unitary operations, to be applied by Charlie and / or Darlie, locally). Therefore distillable entanglement of \( \rho ^{(d + 1)} \) in
the $AC : BD$ cut is at least \( \log_{2}d \) ebit. But here, as
the relative entropy of entanglement of \( \rho ^{(d + 1)} \) in
the $AC : BD$ cut, is less than $\log_{2}d$, so
the distillable entanglement of \( \rho ^{(d + 1)} \), in
the $AC : BD$ cut, should be less than $\log_{2}d$, and hence
a contradiction. Therefore no $(d + 1)$ no. of pairwise orthogonal maximally
entangled states in $d \otimes d$, all taken from the set given in (\ref{psinm}), are distinguishable by LOCC with certainty if
only a single copy of each state is provided. 

What would be the case if we consider local distinguishability of {\it any} $(d + 1)$ no. of pairwise orthogonal maximally entangled states $\left|{\psi}_i^{(max)}\right\rangle$ of $d \otimes d$, instead of considering only states from the set of states given in (\ref{psinm})? Above-mentioned argument will go through if we can extend the incomplete orthogonal basis of maximally entangled states (in $d \otimes d$) $\left\{\left|{\psi}_i^{(max)}\right\rangle : i = 1, 2, \ldots d + 1\right\}$ to a full basis of $d^2$ pairwise orthogonal maximally entangled states of $d \otimes d$. But still now, we don't know whether this extension is possible, in general.

\vspace{0.4cm}
\section{Local distinguishability of maximally entangled states, supplied with two copies}

It has been shown by Horodecki et al. \cite{horo03} that a complete orthonormal basis of $d \otimes d$ can distinguished by LOCC, deterministically or probabilistically, in the single copy case, if and only if all the states are product. Fan \cite{fan03} has shown that the total $d^2$ no. of pairwise orthogonal maximally entangled states $\left|{\Psi}_{nm}^{(d)}\right\rangle$ ($n, m = 0, 1, \ldots, d - 1$), given in (\ref{psinm}), in the single copy case, can never ({\it i.e.}, neither deterministically nor probabilistically) be distinguished by using LOCC only.
    
We are now going to show that any given set of $k$ (where $1 \le k \le d^2$) no. of pairwise orthogonal maximally entangled states in $d \otimes d$, taken from the set given in (\ref{psinm}), can be reliably discriminated by LOCC only, if {\it two} copies of each of these states are provided. To show this we employ the following protocol: We teleport the following state
$|0\rangle$ through the first copy of each of the shared (between Alice and Bob) unknown channel state $\left|{\Psi}_{nm}^{(d)}\right\rangle$, and also we teleport the state $\frac{1}{\sqrt{d}}(|0\rangle + |1\rangle +
\ldots + |d - 1\rangle)$ through the second copy of this shared channel state, by using the standard teleportation protocol ${\cal P}_{00}^{(d)}$ of Bennett et al. \cite{bennett93}, used for each of the above-mentioned two channel states, separately.
Now, after this teleportation
protocol is over, the final two-qudit state at Bob's side, corresponding to two copies of the unknown channel state  $\left|{\Psi}_{nm}^{(d)}\right\rangle$, is given by (modulo a phase):

\begin{widetext}
$$\left(V_{nm}^{(d)} \otimes V_{nm}^{(d)}\right) |0\rangle \otimes \frac{1}{\sqrt{d}}(|0\rangle + |1\rangle + \ldots + |d - 1\rangle) = |m\rangle \otimes \frac{1}{\sqrt{d}} \sum_{j = 0}^{d - 1}~ {\rm exp} \left[\frac{2 \pi i j n}{d}\right] |(j + m)~ {\rm mod}~ d\rangle,$$
\end{widetext}

where $V_{nm}^{(d)}$'s are given in equation (\ref{vdefine}), for $n, m = 0, 1, \ldots, d - 1$. Bob now first does measurement in the computational basis $\{|0\rangle, |1\rangle, \ldots, |d - 1\rangle\}$, on his first qudit. If $|m\rangle$ is the outcome, Bob will then distinguish the following $d$ no. of pairwise orthogonal states  $\frac{1}{\sqrt{d}} \sum_{j = 0}^{d - 1}~ {\rm exp} \left[\frac{2 \pi i j n}{d}\right] |(j + m)~ {\rm mod}~ d\rangle$, where $n = 0, 1, \ldots, d - 1$. And from both the measurement results finally Alice
and Bob will be able to discriminate the $d^2$ no. of pairwise orthogonal
maximally entangled states $\left|{\Psi}_{nm}^{(d)}\right\rangle$, given by equation (\ref{psinm}).

We are unable to proceede in the same way, for local discrimination of {\it any} $d^2$ no. of pairwise orthogonal maximally entangled states in $d \otimes d$, as the most general form of any set of $d^2$ no. of pairwise orthogonal maximally entangled states in $d \otimes d$ is not known \cite{werner}.

\vspace{0.4cm}
\section{Local distinguishability of $d$ no. of maximally entangled states in the single copy case}

Next we discuss the problem of reliable local distinguishability of $d$ no. of pairwise orthogonal maximally entangled states in $d \otimes d$, all taken from the set of states given by equation (\ref{psinm}), in the single copy case. Thus the problem is to test the possibility of reliable local distinguishability of $d$ no. of pairwise orthogonal maximally entangled states $\left|{\Psi}_{n_km_k}^{(d)}\right\rangle$ (in the single copy case), chosen at random from the set of $d^2$ no. of pairwise orthogonal maximally entangled states $\left|{\Psi}_{nm}^{(d)}\right\rangle$, given in (\ref{psinm}), where $n_k, m_k \in \{0, 1, \ldots, d - 1\}$ for $k = 1, 2, \ldots, d$. Without loss of generality, we can assume here that $m_1 \le m_2 \le \ldots \le m_d$. We shall discuss now case-by-case situations. 

\vspace{0.3cm}
{\noindent \underline{\bf {d = 2 :}} According to Walgate et al. \cite{hardy00}, any two pairwise orthogonal maximally entangled states in $2 \otimes 2$ are reliable distinguishable by LOCC, in the single copy case. Let us describe another approach. We shall adopt here the teleportation protocol ${\cal P}_{00}^{(2)}$.} 

{\noindent ({\bf 2.1}) If two far appart parties Alice and Bob share single copies of one of the two states $\left|{\Psi}_{00}^{(2)}\right\rangle$ and $\left|{\Psi}_{10}^{(2)}\right\rangle$, which they want to distinguish with certainty by using LOCC only, Alice will then send the state $\frac{1}{\sqrt{2}}(|0\rangle + |1\rangle)$ through each of these two Bell states using the protocol ${\cal P}_{00}^{(2)}$, and the two output states at Bob's end will be (upto some phases) $V_{00}^{(2)}\frac{1}{\sqrt{2}}(|0\rangle + |1\rangle) = \frac{1}{\sqrt{2}}(|0\rangle + |1\rangle)$ and $V_{10}^{(2)}\frac{1}{\sqrt{2}}(|0\rangle + |1\rangle) = \frac{1}{\sqrt{2}}(|0\rangle - |1\rangle)$
respectively, where $V_{00}^{(2)}$ and $V_{10}^{(2)}$ are given in (\ref{vdefine}). These two states are orthogonal to each other, and hence, they can be reliably distinguished by Bob. So Alice and Bob will be able to tell which of these two Bell states they were sharing initially.} 

{\noindent ({\bf 2.2}) On the other hand, if Alice and Bob share single copy of any one of the following two orthogonal maximally entangled states :
$\left|\Psi_{00}^{(2)}\right\rangle$, $\left|\Psi_{01}^{(2)}\right\rangle$, Alice will then send the state
$|0\rangle$ through each of these two channel states using the
teleportation protocol ${\cal P}_{00}^{(2)}$. Bob will have then the
following two orthogonal states (upto some phases) $V_{00}^{(2)}|0\rangle = |0\rangle$ and $V_{01}^{(2)}|0\rangle = |1\rangle$, which
Bob can easily distinguish, and hence Alice and Bob will come up,
with certainty, which of the above-mentioned two maximally
entangled states they were sharing.} 

Local distinguishability of each of the rest four choices of two no. of pairwise orthogonal maximally entangled states in $2 \otimes 2$ will be either like the case (2.1) or the case (2.2), described above.

Now, {\it any} given set of pairwise orthogonal maximally entangled states of two qubits can be taken as (due to \cite{hardy00}):
\begin{equation} 
\label{mes2}
\begin{array}{lcl}
\left|{\psi}_1\right\rangle &=& \frac{1}{\sqrt{2}}(|00^{\prime}\rangle +~ {\rm exp} [i\theta] |11^{\prime}\rangle), \\ \\
\left|{\psi}_2\right\rangle &=& \frac{1}{\sqrt{2}}(|01^{\prime}\rangle +~ {\rm exp} [i\delta] |10^{\prime}\rangle), 
\end{array}
\end{equation}
where $\{|0\rangle, |1\rangle\}$ is an orthonormal basis on Alice's side, $\{|0^{\prime}\rangle, |1^{\prime}\rangle\}$ is an orthonormal basis on Bob's side, and $\theta, \delta$ are real numbers. We can {\it always} extend this set of two states to the following set $S$ of four pairwise orthogonal maximally entangled states of two qubits
\begin{equation}
\label{genmes2}
\begin{array}{lcl}
\left|{\psi}_1\right\rangle &=& \frac{1}{\sqrt{2}}(|00^{\prime}\rangle +~ {\rm exp} [i\theta] |11^{\prime}\rangle), \\ \\
\left|{\psi}_2\right\rangle &=& \frac{1}{\sqrt{2}}(|01^{\prime}\rangle +~ {\rm exp} [i\delta] |10^{\prime}\rangle), \\ \\
\left|{\psi}_3\right\rangle &=& \frac{1}{\sqrt{2}}(|00^{\prime}\rangle -~ {\rm exp} [i\theta] |11^{\prime}\rangle), \\ \\
\left|{\psi}_4\right\rangle &=& \frac{1}{\sqrt{2}}(|01^{\prime}\rangle -~ {\rm exp} [i\delta] |10^{\prime}\rangle).
\end{array}
\end{equation}
And it is known that \cite{wernerjmp}, corresponding  to the above set $S$, there exists a local unitary operator $U_A \otimes V_B$ ($U_A$ is acting on Alice's system, while $V_B$ is acting on Bob's system), as well as four phases ${\rm exp} \left[i{\theta}_1\right]$,~  ${\rm exp} \left[i{\theta}_2\right]$,~  ${\rm exp} \left[i{\theta}_3\right]$,~  ${\rm exp} \left[i{\theta}_4\right]$, such that 
\begin{equation}
\label{localunitary}
\begin{array}{lcl}
\left|{\psi}_1\right\rangle &=&~ {\rm exp} \left[i{\theta}_1\right] \left(U_A \otimes V_B\right)\left|{\Psi}_{00}^{(2)}\right\rangle, \\ \\ 
\left|{\psi}_2\right\rangle &=&~ {\rm exp} \left[i{\theta}_2\right] \left(U_A \otimes V_B\right)\left|{\Psi}_{01}^{(2)}\right\rangle, \\ \\ 
\left|{\psi}_3\right\rangle &=&~ {\rm exp} \left[i{\theta}_3\right] \left(U_A \otimes V_B\right)\left|{\Psi}_{10}^{(2)}\right\rangle, \\ \\ 
\left|{\psi}_4\right\rangle &=&~ {\rm exp} \left[i{\theta}_4\right] \left(U_A \otimes V_B\right)\left|{\Psi}_{11}^{(2)}\right\rangle. 
\end{array}
\end{equation}
This fact shows that local distinguishability of any two pairwise orthogonal maximally entangled states of two qubits is equivalent to that of any two elements from the set of four states $\left|{\Psi}_{00}^{(2)}\right\rangle$,  $\left|{\Psi}_{01}^{(2)}\right\rangle$,  $\left|{\Psi}_{10}^{(2)}\right\rangle$,  $\left|{\Psi}_{11}^{(2)}\right\rangle$.

\vspace{0.3cm}
{\noindent \underline{\bf d = 3 :} In this case, Walgate et al.'s result \cite{hardy00} is not going to help us directly. Similar to the case for $d = 2$, we shall adopt here the teleportation protocol  ${\cal P}_{00}^{(3)}$.}
 
{\noindent ({\bf 3.1}) If two far appart parties Alice and Bob share single copy of one of the following three mutually orthogonal maximally entangled states in $3 \otimes 3$:
$\left|\Psi_{00}^{(3)}\right\rangle$,
$\left|\Psi_{10}^{(3)}\right\rangle$, and
$\left|\Psi_{01}^{(3)}\right\rangle$, Alice then sends the state $\frac{1}{\sqrt{3}}(|0\rangle +
w|1\rangle + |2\rangle)$ (where $w =~ {\rm exp}\left[\frac{2 \pi
i}{3}\right]$) through each of the three channel states
$\left|\Psi_{00}^{(3)}\right\rangle$,
$\left|\Psi_{10}^{(3)}\right\rangle$ and
$\left|\Psi_{01}^{(3)}\right\rangle$. Bob will have then the following three mutually orthogonal states (upto some phases):
$V_{00}^{(3)}\frac{1}{\sqrt{3}}(|0\rangle + w|1\rangle + |2\rangle) = \frac{1}{\sqrt{3}}(|0\rangle + w|1\rangle + |2\rangle)$,
$V_{10}^{(3)}\frac{1}{\sqrt{3}}(|0\rangle + w|1\rangle + |2\rangle) = \frac{1}{\sqrt{3}}(|0\rangle + w^2|1\rangle + w^2|2\rangle)$, and
$V_{01}^{(3)}\frac{1}{\sqrt{3}}(|0\rangle + w|1\rangle + |2\rangle) = \frac{1}{\sqrt{3}}(|0\rangle + |1\rangle + w|2\rangle)$
respectively. By discriminating these states, Alice and Bob will come up with certainty, which state was they were initially sharing.} 

{\noindent ({\bf 3.2}) If Alice and Bob share single copy of one of the following three pairwise orthogonal maximally entangled states in $3 \otimes 3$: $\left|\Psi_{00}^{(3)}\right\rangle$,
$\left|\Psi_{10}^{(3)}\right\rangle$,
$\left|\Psi_{20}^{(3)}\right\rangle$, then Alice sends the state
$\frac{1}{\sqrt{3}} (|0\rangle + |1\rangle + |2\rangle)$ through each of these three channel states, using ${\cal
P}_{00}^{(3)}$. Bob will have then the following mutually orthogonal
qudits: $V_{00}^{(3)}\frac{1}{\sqrt{3}} (|0\rangle + |1\rangle + |2\rangle) = \frac{1}{\sqrt{3}} (|0\rangle + |1\rangle + |2\rangle)$, $V_{10}^{(3)}\frac{1}{\sqrt{3}} (|0\rangle + |1\rangle + |2\rangle) = \frac{1}{\sqrt{3}} (|0\rangle + w |1\rangle + w^2 |2\rangle)$, $V_{20}^{(3)}\frac{1}{\sqrt{3}} (|0\rangle + |1\rangle + |2\rangle) = \frac{1}{\sqrt{3}} (|0\rangle + w^2
|1\rangle + w^4 |2\rangle)$, respectively. Bob can
discriminate these states reliably, and hence Alice and Bob will
come up with certainty about which of the above-mentioned maximally entangled
states they were sharing.} 

{\noindent ({\bf 3.3}) Consider now the local discrimination
of the following three mutually orthogonal maximally entangled
states in $3 \otimes 3$ (in the single copy case), shared by Alice and Bob: $\left|\Psi_{00}^{(3)}\right\rangle$,
$\left|\Psi_{01}^{(3)}\right\rangle$,
$\left|\Psi_{02}^{(3)}\right\rangle$. If Alice sends the state $|0\rangle$
through each of these three channel states, Bob will have the following mutually orthogonal states (upto some phases):
$|0\rangle$, $|1\rangle$, $|2\rangle$, respectively. Bob can discriminate these states reliably, and
hence Alice and Bob will come up with certainty about which of the above-mentioned three maximally entangled states they were sharing.}  

Each of the rest $^9C_3 - 3 = 81$ choices of three no. of pairwise orthogonal maximally entangled states in $3 \otimes 3$, taken from the set of states given in equation (\ref{psinm}), will be either like the case (3.1) or (3.2) or (3.3), described above. Recently Fan \cite{fan03} has also shown that any three states taken from equation (\ref{psinm}), for $d = 3$, can be reliably locally distinguished.  

If Alice and Bob, instead, share one of the following three pairwise orthogonal maximally entangled states of $3 \otimes 3$ (in the single copy case) 
$$\left|{\psi}_i^{(3)}\right\rangle = \left(I \otimes V_i\right)\left|{\Psi}_{00}^{(3)}\right\rangle,~~ i = 1, 2, 3,$$
and if they use the teleportation protocol  ${\cal P}_{00}^{(3)}$ to teleport a state $\left|{\phi}^{(3)}\right\rangle$ (from Alice's place to Bob's place), we have seen earlier that the output states at Bob's side, for the occurance of Alice's measurement result as $\left|{\Psi}_{nm}^{(3)}\right\rangle = \left(U_{nm}^{(3)} \otimes I\right)\left|{\Psi}_{00}^{(3)}\right\rangle$ (where $n, m = 0, 1, 2$), will be $\left(U_{nm}^{(3)}V_i\left(U_{nm}^{(3)}\right)^{\dagger}\right)\left|{\phi}^{(3)}\right\rangle$, for $i = 1, 2, 3$. And as in the above-mentioned cases (3.1) - (3.3), we would like to get these three (for $i = 1, 2, 3$) states to be orthogonal to each other, for each values of the pair $(n, m)$, {\it i.e.}, we demand that for each choice of the pair $n, m \in \{0, 1, 2\}$, we must have
\begin{equation}
\label{3dcondition}
\left\langle{\phi}^{(3)}\right|U_{nm}^{(3)}V_j^{\dagger}V_i\left(U_{nm}^{(3)}\right)^{\dagger}\left|{\phi}^{(3)}\right\rangle = 0,
\end{equation} 
for $i \ne j$ and $i, j = 1, 2, 3$. Even though here ${\rm Tr} \left(V_j^{\dagger}V_i\right) = 3{\delta}_{ij}$ (which is equivalent to the condition for orthogonality of the maximally entangled states $\left|{\psi}_j^{(3)}\right\rangle$ and $\left|{\psi}_i^{(3)}\right\rangle$), conditions in (\ref{3dcondition}) {\it do not} hold good, in the most general situation. On the other hand, it is not known whether conditions like (\ref{localunitary}) also hold good for $d = 3$. Thus the question of reliable local distinguishability of {\it any} three pairwise orthogonal maximally entangled states in $3 \otimes 3$, in the single copy case, and in the most general situation, is not yet known.   

\vspace{0.3cm}
{\noindent {\underline{\bf Some general cases :}} It is easy to verify that both the cases (3.2) and (3.3) can be extended to respective $d$-dimensional situations. In fact, the following set of $d$ no. of pairwise orthogonal maximally entangled states:  $\left|\Psi_{0m}^{(d)}\right\rangle$, $\left|\Psi_{1m}^{(d)}\right\rangle$, $\ldots$, $\left|\Psi_{(d - 1)m}^{(d)}\right\rangle$ can be reliably discriminated by LOCC, in the single copy case, for any given value of $m$ from the set $\{0, 1, \ldots, d - 1\}$. This can be achieved by sending the state $\frac{1}{\sqrt{d}} (|0\rangle + |1\rangle + \ldots + |d - 1\rangle)$ through each of these $d$ channel states, using ${\cal
P}_{00}^{(d)}$, as the corresponding $d$ no. of output states $V_{nm}^{(d)}\frac{1}{\sqrt{d}} (|0\rangle + |1\rangle + \ldots + |d - 1\rangle) = \frac{1}{\sqrt{d}} \sum_{j = 0}^{d - 1}~ {\rm exp} \left[\frac{2 \pi i j n}{d}\right] |(j + m)~ {\rm mod}~ d\rangle$ (for $n = 0, 1, 2, \ldots, d - 1$) are pairwise orthogonal. Similarly the following set of $d$ no. of pairwise orthogonal maximally entangled states:  $\left|\Psi_{n0}^{(d)}\right\rangle$, $\left|\Psi_{n1}^{(d)}\right\rangle$, $\ldots$, $\left|\Psi_{n (d - 1)}^{(d)}\right\rangle$ can be reliably discriminated by LOCC, in the single copy case, for any given value of $n$ from the set $\{0, 1, \ldots, d - 1\}$. This can be achieved by sending the state $|0\rangle$ through each of these $d$ channel states, using ${\cal
P}_{00}^{(d)}$, as the corresponding $d$ no. of output states $V_{nm}^{(d)}|0\rangle = |m\rangle$ (for $m = 0, 1, 2, \ldots, d - 1$) are orthogonal to each other.}

\vspace{0.3cm}
{\noindent {\underline{\bf d = 4 :}} Here we will use the teleportation protocol ${\cal P}_{00}^{(4)}$. Appart from the results, which are valid for general $d$, there are some cases where four pairwise orthogonal maximally entangled states in $4 \otimes 4$ can be reliably distinguished (in the single copy case) by LOCC, using the teleportation protocol ${\cal P}_{00}^{(4)}$. But there are exceptions also.}

{\noindent ({\bf 4.1}) Let two far apart parties Alice and Bob share single copy of one of the following four pairwise orthogonal maximally entangled states of $4 \otimes 4$: $\left|\Psi_{00}^{(4)}\right\rangle$, $\left|\Psi_{10}^{(4)}\right\rangle$, $\left|\Psi_{0m}^{(4)}\right\rangle$ and $\left|\Psi_{1m}^{(4)}\right\rangle$, where $m$ is any one of the following three values $1, 2, 3$. For $m = 1$ case, Alice will have to teleport the state $\frac{1}{2}(|0\rangle + i|1\rangle + |2\rangle + i|3\rangle)$ through each of the above-mentioned four channel states, for $m = 2$, Alice will have to teleport the state $\frac{1}{2}(|0\rangle + |1\rangle - |2\rangle + |3\rangle)$ through each of the corresponding above-mentioned four channel states, and for $m = 3$, Alice will have to teleport the state $\frac{1}{2}(|0\rangle - i|1\rangle + |2\rangle - i|3\rangle)$ through each of the corresponding above-mentioned four channel states. In each of these three cases, the four final states at Bob's hand will be pairwise orthogonal, and hene reliable local discrimination of the corresponding four maximally entangled states is possible.}

In a similar way, it can be shown that the following four states can be reliably distinguished by LOCC, in the single copy case: $\left|\Psi_{n_1m_1}^{(4)}\right\rangle$, $\left|\Psi_{n_2m_1}^{(4)}\right\rangle$, $\left|\Psi_{n_1m_2}^{(4)}\right\rangle$, $\left|\Psi_{n_2m_2}^{(4)}\right\rangle$, where $n_1 \ne n_2$, $m_1 \ne m_2$, and $n_1, n_2, m_1, m_2 \in \{0, 1, 2, 3\}$.

{\noindent ({\bf 4.2)} Above-mentioned programme of reliable local discrimination by teleportation fails in a case when Alice and Bob share single copy of one of the following four states: $\left|\Psi_{00}^{(4)}\right\rangle$, $\left|\Psi_{10}^{(4)}\right\rangle$, $\left|\Psi_{20}^{(4)}\right\rangle$ and $\left|\Psi_{02}^{(4)}\right\rangle$. This happens because there exists {\it no} input state $\left|{\phi}^{(4)}\right\rangle$ for which $V_{00}^{(4)}\left|{\phi}^{(4)}\right\rangle$, $V_{10}^{(4)}\left|{\phi}^{(4)}\right\rangle$, $V_{20}^{(4)}\left|{\phi}^{(4)}\right\rangle$, $V_{02}^{(4)}\left|{\phi}^{(4)}\right\rangle$ are pairwise orthogonal. This failurity does not depend on the choice of the teleportation protocol. In fact, we would like to mention here that if we allow {\it only} one--way protocols for discriminating the above-mentioned four states, the reliable discrimination of these four states will be then {\it impossible}  \cite{ft2}. So there will be {\it no} local basis transformation via which these four states can be rewritten as 
\begin{equation}
\label{wshvform}
\begin{array}{lcl} 
\left|\Psi_{00}^{(4)}\right\rangle &=& \displaystyle{\left|0^{\prime}{\alpha}_1\right\rangle + \left|1^{\prime}{\beta}_1\right\rangle + \left|2^{\prime}{\gamma}_1\right\rangle + \left|3^{\prime}{\delta}_1\right\rangle},\\ \\
\left|\Psi_{10}^{(4)}\right\rangle &=& \displaystyle{\left|0^{\prime}{\alpha}_2\right\rangle + \left|1^{\prime}{\beta}_2\right\rangle + \left|2^{\prime}{\gamma}_2\right\rangle + \left|3^{\prime}{\delta}_2\right\rangle},\\ \\
\left|\Psi_{20}^{(4)}\right\rangle &=& \displaystyle{\left|0^{\prime}{\alpha}_3\right\rangle + \left|1^{\prime}{\beta}_3\right\rangle + \left|2^{\prime}{\gamma}_3\right\rangle + \left|3^{\prime}{\delta}_3\right\rangle},\\ \\ 
\left|\Psi_{02}^{(4)}\right\rangle &=& \displaystyle{\left|0^{\prime}{\alpha}_4\right\rangle + \left|1^{\prime}{\beta}_4\right\rangle + \left|2^{\prime}{\gamma}_4\right\rangle + \left|3^{\prime}{\delta}_4\right\rangle},
\end{array}
\end{equation}
 where $\left|0^{\prime}\right\rangle$, $\left|1^{\prime}\right\rangle$, $\left|2^{\prime}\right\rangle$, and $\left|3^{\prime}\right\rangle$ are pairwise orthogonal states of a four dimensional Hilbert space, and $\left\langle\alpha_i|\alpha_j\right\rangle = \left\langle\beta_i|\beta_j\right\rangle = \left\langle\gamma_i|\gamma_j\right\rangle = \left\langle\delta_i|\delta_j\right\rangle = 0$ if $i \ne j$ -- a sufficient condition for reliable local discrimination \cite{hardy00}.}      
    
$\left\{\left|\Psi_{10}^{(4)}\right\rangle, \left|\Psi_{20}^{(4)}\right\rangle, \left|\Psi_{30}^{(4)}\right\rangle, \left|\Psi_{12}^{(4)}\right\rangle\right\}$ is another set of four locally indistinguishable states (by the teleportation protocol, described above) pairwise orthogonal maximally entangled states in $4 \otimes 4$, like the one described in (4.2).

\vspace{0.3cm}
{\noindent {\underline{\bf d = 5 and beyond:}} For the case when $d = 5$, there are non-trivial ({\it i.e.}, sets of states which are not of the two general forms, described above, for any $d$) sets of five states, taken from the set of states given in equation (\ref{psinm}), which can be reliably distinguished by using the teleportation protocol. One such example is the set of following five states: $\left|{\Psi}_{00}^{(5)}\right\rangle$, $\left|{\Psi}_{10}^{(5)}\right\rangle$, $\left|{\Psi}_{20}^{(5)}\right\rangle$, $\left|{\Psi}_{30}^{(5)}\right\rangle$, $\left|{\Psi}_{03}^{(5)}\right\rangle$. On the other hand, just like the case (4.2), there are sets of states (e.g., $\left|{\Psi}_{00}^{(5)}\right\rangle$, $\left|{\Psi}_{11}^{(5)}\right\rangle$, $\left|{\Psi}_{21}^{(5)}\right\rangle$, $\left|{\Psi}_{13}^{(5)}\right\rangle$, $\left|{\Psi}_{23}^{(5)}\right\rangle$) which can not be reliably distinguished by using the teleportation protocol ${\cal P}_{00}^{(5)}$. Similarly, for $d = 6$, $\left\{\left|\Psi_{00}^{(6)}\right\rangle, \left|\Psi_{10}^{(6)}\right\rangle, \left|\Psi_{20}^{(6)}\right\rangle, \left|\Psi_{30}^{(6)}\right\rangle, \left|\Psi_{40}^{(6)}\right\rangle, \left|\Psi_{03}^{(6)}\right\rangle,\right\}$ is a set of six pairwise orthogonal maximally entangled states in $6 \otimes 6$, which can not be reliably distinguished (in the single copy case), by the above-mentioned teleportation method.}    

Above discussions give rise to the following sufficient condition for reliable local distinguishability of sets of maximally entangled states, taken from the set given by equation (\ref{psinm}).

\vspace{0.3cm}

{\noindent {\underline{\bf Local distingushability of less than d states :}} Fan \cite{fan03} has shown that any $l$ no. of pairwise orthogonal maximally entangled states of $d \otimes d$, taken from equation (\ref{psinm}), can be reliably distinguished by LOCC if $l(l-1) \le 2d$. This shows that any three states in $5 \otimes 5$, taken from equation (\ref{psinm}), can be distinguished. Then there is a possibility of finding a set of four states in $5 \otimes 5$ which cannot be locally discriminable. We provide one such possible example which is the set consisting of $\left|{\Psi}_{11}^{(5)}\right\rangle$, $\left|{\Psi}_{21}^{(5)}\right\rangle$, $\left|{\Psi}_{13}^{(5)}\right\rangle$, $\left|{\Psi}_{23}^{(5)}\right\rangle$ (given in Eq.(\ref{psinm})), which does not satisfy the conjectured necessary condition for distinguishability (given in the next section)}.     

\vspace{0.2cm}

{\noindent {\bf Sufficient condition for distinguishability of $\left|{\Psi}_{nm}^{(d)}\right\rangle$'s:} {\it Single copies of $L$ no. of pairwise orthogonal maximally entangled states $\left|{\Psi}_{n_im_i}^{(d)}\right\rangle$ (for $i = 1, 2, \ldots, L$), taken from the set given in equation (\ref{psinm}), can be reliably discriminated by LOCC} 
{\it if there exists at least one} 
{\it state $\left|{\phi}^{(d)}\right\rangle$ for which the states $V_{n_1m_1}^{(d)}\left|{\phi}^{(d)}\right\rangle$,  $V_{n_2m_2}^{(d)}\left|{\phi}^{(d)}\right\rangle$, $\ldots$,  $V_{n_Lm_L}^{(d)}\left|{\phi}^{(d)}\right\rangle$ are pairwise orthogonal, where $V_{nm}^{(d)}$'s are given by equation (\ref{vdefine}).}}

\vspace{0.4cm}
\section{Necessary condition for reliable distinguishability}

For the sets of maximally entangled states, given in (4.2), or in examples for cases where $d \ge 5$, described above (where the local discrimination by teleportation failed), we have seen that the above-mentioned sufficient condition is not satisfied. Does it mean that none of these sets of states can be reliably discriminated by LOCC only, in the single copy case?  Or ({\it i.e.}, contrapositively), we want to check whether {\it reliable local distinguishability of single copies (or, multiple copies) of $L$ no. of pairwise orthogonal maximally entangled states  $\left|{\Psi}_{n_im_i}^{(d)}\right\rangle$ (for $i = 1, 2, \ldots, L$) implies the existence of at least one state $\left|{\phi}^{(d)}\right\rangle$ for which the states $V_{n_1m_1}^{(d)}\left|{\phi}^{(d)}\right\rangle$,  $V_{n_2m_2}^{(d)}\left|{\phi}^{(d)}\right\rangle$, $\ldots$,  $V_{n_Lm_L}^{(d)}\left|{\phi}^{(d)}\right\rangle$ are pairwise orthogonal}. Later on, we shall discuss about this implication (written in italics), when it is true, as a {\it necessary} condition of local distinguishability.

{\noindent {\bf (i) Distinguishability of any two orthogonal maximally entangled states:} Let $\left|{\psi}_1^{(d)}\right\rangle$ and $\left|{\psi}_2^{(d)}\right\rangle$ be given any two pairwise orthogonal maximally entangled states of $d \otimes d$. These two states are reliably distinguishable by LOCC only \cite{hardy00}. For these two states, we can always find an orthonormal basis $\{|0\rangle, |1\rangle, \ldots, |d - 1\rangle\}$ for Alice's system, and another orthonormal basis $\{|0^{\prime}\rangle, |1^{\prime}\rangle, \ldots, |(d - 1)^{\prime}\rangle\}$ for Bob's system such that 

\begin{widetext}
\begin{equation}
\label{2necessary}
\begin{array}{lclcl}
\left|{\psi}_1^{(d)}\right\rangle &=& (I \otimes I)\left|{\psi}_1^{(d)}\right\rangle &=& \frac{1}{\sqrt{d}}\left(|00^{\prime}\rangle +~ {\rm exp} \left[i{\theta}_1\right] |11^{\prime}\rangle + \ldots +~ {\rm exp} \left[i{\theta}_{d - 1}\right] |(d - 1)(d - 1)^{\prime}\rangle\right), \\ \\ 
\left|{\psi}_2^{(d)}\right\rangle &=& (I \otimes V)\left|{\psi}_1^{(d)}\right\rangle &=& \frac{1}{\sqrt{d}}\left(|00^{\prime \prime}\rangle +~ {\rm exp} \left[i{\delta}_1\right] |11^{\prime \prime}\rangle + \ldots +~ {\rm exp} \left[i{\delta}_{d - 1}\right] |(d - 1)(d - 1)^{\prime \prime}\rangle\right),
\end{array}
\end{equation}
\end{widetext}

where $\{|0^{\prime \prime}\rangle, |1^{\prime \prime}\rangle, \ldots, |(d - 1)^{\prime \prime}\rangle\}$ is an orthonormal basis of Bob's system, $\langle j^{\prime} | j^{\prime \prime} \rangle = 0$ for $j = 0, 1, 2, \ldots, d - 1$, and $V$ is an unitary operator, acting on Bob's system, such that $V|0^{\prime}\rangle = |0^{\prime \prime}\rangle$,  $V|1^{\prime}\rangle =~ {\rm exp} \left[i{\delta}_1\right]|1^{\prime \prime}\rangle$, $\ldots$, $V|(d - 1)^{\prime}\rangle =~ {\rm exp} \left[i{\delta}_{d - 1}\right]|(d - 1)^{\prime \prime}\rangle$ \cite{hardy00}. Thus there are $d$ states $|0^{\prime}\rangle$, $|1^{\prime}\rangle$, $\ldots$, $|(d - 1)^{\prime}\rangle$ (of Bob's system) for which $\{I|0^{\prime}\rangle, V|0^{\prime}\rangle\}$, $\{I|1^{\prime}\rangle, V|1^{\prime}\rangle\}$, $\ldots$, $\{I|(d - 1)^{\prime}\rangle, V|(d - 1)^{\prime}\rangle\}$ are $d$ pairs of orthogonal states.}                  

\vspace{0.2cm}
{\noindent {\bf (ii) Distinguishability of all $\left|{\Psi}_{nm}^{(d)}\right\rangle$'s when two copies are given :} We have seen that the $d^2$ no. of pairwise orthogonal maximally entangled states, given by equation (\ref{psinm}), are reliably distinguishable by using LOCC only, if two copies of each of these states are given. Let 
$$\left|{\chi}_{nm}\right\rangle_{AC : BD} = \left|{\Psi}_{nm}^{(d)}\right\rangle_{AB} \otimes \left|{\Psi}_{nm}^{(d)}\right\rangle_{CD},$$ 
for $n, m = 0, 1, \ldots, d - 1$ and where Alice possesses the two systems $A$, $C$, and Bob possesses the other two systems $B$, $D$. Thus we see that when Alice and Bob share single copy of one of the $d^2$ no. of pairwise orthogonal maximally entangled states  $\left|{\chi}_{nm}\right\rangle_{AC : BD}$ of $d^2 \otimes d^2$, they can reliably distinguish these states, using LOCC only. Also here $$\left(I^{AC} \otimes W_{nm}^{BD}\right)\left|{\chi}_{nm}\right\rangle_{AC : BD} = \left|{\chi}_{00}\right\rangle_{AC : BD},$$
where $I^{AC}$ is the identity operator on the $d^2$-dimensional Hilbert space of Alice, while $W_{nm}^{BD} = V_{nm}^{(d)} \otimes V_{nm}^{(d)}$ is an unitary operator acting on the $d^2$-dimensional Hilbert space of Bob, where $V_{nm}^{(d)}$ is given by (\ref{vdefine}). Let us consider the state $\left|{\phi}^{(d^2)}\right\rangle = |0\rangle \otimes \frac{1}{\sqrt{d}}(|0\rangle + |1\rangle + \dots + |d - 1\rangle)$. It can be shown that (infact, we have shown it earlier, in this paper) the states $W_{nm}\left|{\phi}^{(d^2)}\right\rangle$ (for $n, m = 0, 1, \ldots, d - 1$) are pairwise orthogonal.}   

\vspace{0.2cm}
For each of sets of $d$ pairwise orthogonal maximally entangled states $\left|{\Psi}_{n_im_i}^{(d)}\right\rangle$ (for $i = 1, 2, \ldots, d$), discussed earlier for particular values of $d$, where the states (of the set) can be shown to be reliably distinguishable (using suitable teleportation protocols) by LOCC only, we are able to find an input state $\left|{\phi}^{(d)}\right\rangle$ such that the following $d$ no. of states $V_{n_im_i}^{(d)}\left|{\phi}^{(d)}\right\rangle$ (for $i = 1, 2, \ldots, d$) are orthogonal to each other. On the other hand, for particular values of $d$, we have seen that there are examples of sets of $d$ states  $\left|{\Psi}_{n_im_i}^{(d)}\right\rangle$ (for $i = 1, 2, \ldots, d$), where one can never find a pure state $\left|{\phi}^{(d)}\right\rangle$ such that the $d$ no. of states  $V_{n_im_i}^{(d)}\left|{\phi}^{(d)}\right\rangle$ (for $i = 1, 2, \ldots, d$) are orthogonal to each other. And in each of these examples, the states are possibly reliably locally indistinguishable (e.g. the four pairwise orthogonal maximally entangled states $\left|{\Psi}_{00}^{(4)}\right\rangle$,  $\left|{\Psi}_{10}^{(4)}\right\rangle$,  $\left|{\Psi}_{20}^{(4)}\right\rangle$,  $\left|{\Psi}_{02}^{(4)}\right\rangle$ of $4 \otimes 4$ can be shown to be reliably indistinguishable by using 1--way LOCC only, and there exists no state $\left|{\phi}^{(4)}\right\rangle$, for which the four states $V_{00}^{(4)}\left|{\phi}^{(4)}\right\rangle$,  $V_{10}^{(4)}\left|{\phi}^{(4)}\right\rangle$,  $V_{20}^{(4)}\left|{\phi}^{(4)}\right\rangle$,  $V_{02}^{(4)}\left|{\phi}^{(4)}\right\rangle$, are orthogonal to each other). All these facts lead to the following conjecture, in terms of the above-mentioned necessary condition:

\vspace{0.2cm}
{\noindent {\bf Conjecture :} {\it Let $S = \left\{\left|{\Phi}_i^{(d)}\right\rangle = \left(I \otimes V_i^{(d)}\right)\left|{\Psi}_{00}^{(d)}\right\rangle~ :~ i = 1, 2, \ldots, L\right\}$ be any given set of $L$ no. of pairwise orthogonal maximally entangled states of $d \otimes d$, where $V_i^{(d)}$'s are unitary operators acting on the states of a $d$-dimensional Hilbert space, and $2 \le L \le d^2$. If these $L$ no. of states are reliably distinguishable by using LOCC only, then there will always exist at least one state $\left|{\phi}^{(d)}\right\rangle$ of the $d$-dimensional Hilbert space, for which the $L$ states $V_i^{(d)}\left|{\phi}^{(d)}\right\rangle$ (for $i = 1, 2, \ldots, L$) are pairwise orthogonal.}}

\vspace{0.4cm}
\section{Conclusion}

In conclusion, we have shown that more than $d$ no. of pairwise orthogonal maximally entangled states in $d \otimes d$, all taken from the set given in (\ref{psinm}), can not be reliably discriminated, in the single copy case, by using LOCC only, but they can be reliably discriminated, by using LOCC only, if two copies of each of the states are given. It has been shown here, using the standard teleportation protocol of Bennett et al. \cite{bennett93}, that for $d \le 3$, any $d$ no. of pairwise orthogonal maximally entangled states in $d \otimes d$ can be reliably discriminated, in the single copy case, by using LOCC only, when all the states are taken from the set given in (\ref{psinm}) - the same result has also been obtained by Fan \cite{fan03}. But for $d \ge 4$, our method of discrimination, by using the above-mentioned standard teleportation protocol, fails in some cases, and we are undecisive in this situation, regarding reliable local distinguishability of $d$ no. of pairwise orthogonal maximally entangled states in $d \otimes d$, all taken from the set given in (\ref{psinm}). Whether the most general type of $d$ or less than $d$ no. of pairwise orthogonal maximally entangled states of $d \otimes d$ ({\it i.e.}, maximally entangled states which are not necessarily of the form of equation (\ref{psinm})) are reliably locally distinguishable, in the single copy case, is yet to be settled fully. Fan \cite{fan03} provided a partial answer to this question.  

If the above-mentioned conjecture is true, one can easily see that {\it no} $L$ no. of pairwise orthogonal maximally entangled states $\left|{\Phi}_i^{(d)}\right\rangle \equiv \left(I \otimes V_i^{(d)}\right)\left|{\Psi}_{00}^{(d)}\right\rangle$ (for $i = 1, 2, \ldots, L$) of $d \otimes d$ can be reliably discriminated by using LOCC only, and in the single copy case, if $L \ge (d + 1)$. This is so, because there would be no room for the existence of $L$ no. of pairwise orthogonal states $V_i^{(d)}\left|{\phi}^{(d)}\right\rangle$ (for $i = 1, 2, \ldots, L$) in a $d$-dimensional Hilbert space, if $L \ge (d + 1)$. It is to be noted here that the maximally entangled states  $\left|{\Phi}_i^{(d)}\right\rangle$ are not necessarily of the form, given in equation (\ref{psinm}).

While giving the sufficient condition for reliable local discrimination of pairwise orthogonal maximally entangled states, we restricted ourselves to states which are of the form, given in equation (\ref{psinm}). This is so because there are examples of sets of pairwise orthogonal maximally entangled states, for which the above-mentioned necessary condition (given by the conjecture) is satisfied ({\it i.e.}, one can find at least one state $\left|{\phi}^{(d)}\right\rangle$, for which the states $V_i^{(d)}\left|{\phi}^{(d)}\right\rangle$ are pairwise orthogonal), but local discrimination, by using standard teleportation protocol, fails. One such example is the following set of three pairwise orthogonal maximally entangled states of $3 \otimes 3$:
$$\left|{\psi}_1\right\rangle = \frac{1}{\sqrt{3}}(|00\rangle + |11\rangle + |22\rangle),$$
$$\left|{\psi}_2\right\rangle = \frac{1}{\sqrt{3}}(|01\rangle + |12\rangle + |20\rangle),$$
$$\left|{\psi}_3\right\rangle = \frac{1}{\sqrt{3}}\left(\left|0{\phi}_0\right\rangle + \left|1{\phi}_1\right\rangle + \left|2{\phi}_2\right\rangle\right),$$
where $\left|{\phi}_0\right\rangle = \frac{1}{\sqrt{3}}(|0\rangle + |1\rangle + |2\rangle)$,  $\left|{\phi}_1\right\rangle = \frac{1}{\sqrt{3}}(|0\rangle + \omega|1\rangle + {\omega}^2|2\rangle)$,  $\left|{\phi}_0\right\rangle = \frac{1}{\sqrt{3}}(|0\rangle + {\omega}^2|1\rangle + {\omega}|2\rangle)$, and where $\omega =~ {\rm exp} \left[\frac{2 \pi i}{3}\right]$. Although failure of local discrimination by using the standard teleportation protocol does not guarantee the same for all other teleportation protocols, we are, still now, unable to reliably distinguish the above-mentioned three pairwise orthogonal maximally entangled states in $3 \otimes 3$, by using any teleportation protocol.

\section*{}

\vspace{0.2cm}
{\noindent \large{\bf Acknowledgement}}

Authors thank Arun K. Pati for useful discussion of the problem
and  Somshubhro Bandyopadhyay for earlier discussion on
general aspect of local discrimination via teleportation. Authors also thank Samir Kunkri for his help in calculation of the problem of distinguishability of some classes of maximally entangled states discussed here. Part of the work was done when S. G. was working in Indian Statistical Institute, Kolkata, as a Post-doctoral fellow of National Board of Higher Mathematics, Department of Atomic Energy, Govt. of India.


\vspace{0.4cm}
\begin{thebibliography}{99}
\bibitem{bennett99} C. H. Bennett, D. P. DiVincenzo, C. A. Fuchs, T. Mor, E. Rains,
P. W. Shor, J. A. Smolin, and W. K. Wootters, \emph{Phys. Rev. A}
\textbf{59} (1999) 1070 
\bibitem{hardy00} J. Walgate, A. J. Short, L. Hardy and V. Vedral, \emph{Phys. Rev.
Lett.} \textbf{85} (2000) 4972 
\bibitem{hardy02} J. Walgate and L. Hardy, {\it Phys. Rev. Lett.} {\bf 89} (2002) 147901
\bibitem{ghosh01} S. Ghosh, G. Kar, A. Roy, A. Sen (De) and U. Sen,
\emph{Phys. Rev. Lett.} \textbf{87} (2001) 277902
\bibitem{fan03} H. Fan, \emph{quant-ph/0311026} 
\bibitem{braunstein} S. L. Braunstein, G. M. D'Ariano, G. J. Milburn, and M. F. Sacchi, {\it Phys. Rev. Lett.} {\bf 84} (2000) 3486
\bibitem{vedral97} V. Vedral, M. B. Plenio, M. A. Rippin and P. L. Knight, \emph{Phys. Rev. Lett.} \textbf{78} (1997) 2275 
; V. Vedral
and M. B. Plenio, \emph{Phys. Rev. A} \textbf{57} (1998) 1619 
\bibitem{rains99} E. M. Rains, \emph{Phys. Rev. A} \textbf{60} (1999) 179 
\bibitem{horo00} M. Horodecki, P. Horodecki and R. Horodecki, \emph{Phys. Rev. Lett.}
\textbf{84} (2000) 2014 
\bibitem{horo03} M. Horodecki, A. Sen (De), U. Sen and K. Horodecki, \emph{Phys. Rev. Lwtt.} \textbf{90} (2003) 047902
\bibitem{bennett93} C. H. Bennett, G. Brassard, C. Cr\'{e}peau, R. Josza, A. Peres
and W. K. Wootters, Phys. Rev. Lett. {\bf 70} (1993) 1895
\bibitem{werner} R. F. Werner, {\it J. Phys. A : Math. Gen.} {\bf 34} (2001) 7081 
\bibitem{wernerjmp} K. G. H. Vollbrecht and R. F. Werner, {\it J. Math. Phys.} {\bf 41} (2000) 6772 
\bibitem{ft2} Thus, we here assume that either Alice or Bob starts (say, Alice, {\i.e.}, ``Alice going first'' \cite{hardy02}) doing generalized measurements $\left\{A_i\right\}_{i = 1}^{N}$, where $\sum_{i = 1}^{N} A_i^{\dagger}A_i = I$, and the resulting states $Tr_A\left[\left(A_i \otimes I\right)P\left[\left|\Psi_{00}^{(4)}\right\rangle\right]\left(A_i^{\dagger} \otimes I\right)\right]$, $Tr_A\left[\left(A_i \otimes I\right)P\left[\left|\Psi_{10}^{(4)}\right\rangle\right]\left(A_i^{\dagger} \otimes I\right)\right]$, $Tr_A\left[\left(A_i \otimes I\right)P\left[\left|\Psi_{20}^{(4)}\right\rangle\right]\left(A_i^{\dagger} \otimes I\right)\right]$ and  $Tr_A\left[\left(A_i \otimes I\right)P\left[\left|\Psi_{02}^{(4)}\right\rangle\right]\left(A_i^{\dagger} \otimes I\right)\right]$ at Bob's end (corresponding to each measurement outcome `$i$' of Alice's generalized measurement) will be pairwise orthogonal (including the case when one or more of these resulting states becomes a null state), so that Bob can then reliably distinguish these pairwise orthogonal states, and hence the discrimination protocol is over -- {\it no} further operation has to be done by Alice or Bob, on their respective subsystems. Now choosing $A_i = \sum_{j = 0}^{3} \left|{\phi}_{ij} \right\rangle \left\langle j\right|$ (for $i = 1, 2, \ldots, N$), where $\left|{\phi}_{i0} \right\rangle$, $\left|{\phi}_{i1} \right\rangle$, $\left|{\phi}_{i2} \right\rangle$, $\left|{\phi}_{i3} \right\rangle$ are states of some $d$-dimensioanl Hilbert space, and they are not necessarily normalized, not necessarily orthogonal to each other, but $\sum_{i = 1}^{N} \left\langle{\phi}_{ij} | {\phi}_{ik}\right\rangle = {\delta}_{jk}$, for $j, k = 0, 1, 2, 3$, one can verify that the above-mentioned orthogonality conditions will always give rise to a contradiction.  

\end{thebibliography}
\end{document}